\documentclass[twocolumn,prb,showpacs,citeautoscript,floatfix]{revtex4}
\usepackage{graphicx}
\input{epsf}

\def\full{C$_{60}$}
\def\PhC60{(Ph$_{4}$P)$_{2}$IC$_{60}$}
\def\cm-1{cm$^{-1}$}
\def\C60a{C$_{60}^{-}$}

\begin{document}
\title{Pressure-induced transition from the dynamic to static Jahn-Teller effect \\
in (Ph$_{4}$P)$_{2}$IC$_{60}$}
\author{E. A. Francis$^{1}$, S. Scharinger$^{1}$, K. N\'{e}meth$^{2}$,
K. Kamar\'{a}s$^{2}$}
\author{C. A. Kuntscher$^{1}$}
\email[E-mail:~]{christine.kuntscher@physik.uni-augsburg.de}
\affiliation{$^{1}$Experimentalphysik 2, Universit\"at Augsburg, D-86135
Augsburg, Germany}
\affiliation{$^{2}$Institute for Solid State Physics and Optics,
Wigner Research Centre for Physics, Hungarian Academy of Sciences, P. O. Box 49, Budapest, Hungary, H-1525}
\date{\today}

\begin{abstract}
High-pressure infrared transmission measurements on \PhC60 were performed up to 9 GPa over a
broad frequency range (200 - 20000 cm$^{-1}$) to monitor the vibrational and
electronic/vibronic excitations under pressure. The four fundamental T$_{1u}$ modes of \C60a\
are split into doublets already at the lowest applied pressure and harden with increasing pressure.
Several cation modes and fullerene-related modes split into doublets at around 2~GPa, the most prominent
one being the G$_{1u}$ mode. The splitting of the vibrational modes can be attributed to the transition
from the dynamic to static Jahn-Teller effect, caused by steric crowding at high pressure.
Four absorption bands are observed in the NIR-VIS frequency
range. They are discussed in terms of transitions between LUMO electronic states in \C60a, which are split because of the
Jahn-Teller distortion and can be coupled with vibrational
modes. Various distortions and the corresponding symmetry lowering are discussed.
The observed redshift of the absorption bands indicates that the splitting of the LUMO electronic states
is reduced upon pressure application.
\end{abstract}

\maketitle

\section{Introduction}

\begin{figure}[t]   
\begin{center}
\includegraphics*[scale=0.65]{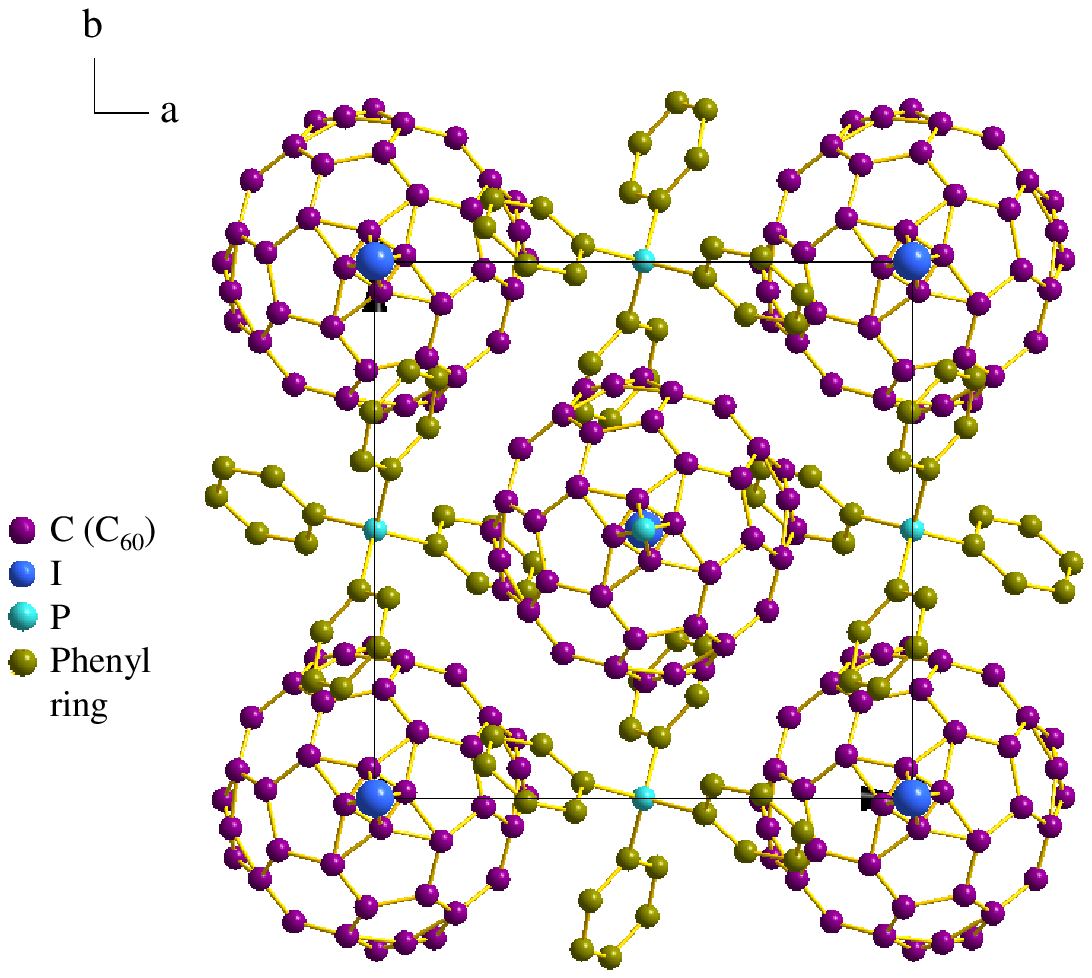}
\caption{Crystal structure of \PhC60.\cite{Penicaud93, Bilow95, Pilz02} } \label{crystal}
\end{center}
\end{figure}

Tetraphenylphosphonium iodide-\full\ [\PhC60] is a prototype to
study \C60a\ radical anion in a solid state environment, as large
cations (Ph$_{4}$P)$_{2}$I are so well positioned that they separate the \full\ ions from each other (see Fig.~\ref{crystal}).
The weak coupling between the \C60a\ and the (Ph$_{4}$P$^{+}$)X$^{-}$ structural units
was demonstrated by Raman measurements on (Ph$_{4}$P)$_{2}$C$_{60}$X (X=Cl, Br, I) showing the
insensitivity of the spectra to the halogen anion.\cite{Sauvajol97}
The compounds (Ph$_{4}$P)$_{2}$C$_{60}$X have the advantage of being air stable unlike the
other fullerides. Also they can be grown as single crystals in contrast to the powder form of several other fullerides.
\PhC60\ crystallizes in a tetragonal structure with the space group I4/m.\cite{Penicaud93, Bilow95, Pilz02} At room temperature,
the dynamic nature of the Jahn-Teller (JT) effect in (Ph$_{4}$Y)$_{2}$XC$_{60}$ (\textit{Y} =
P or As, \textit{X} =
Cl, I, or Br) was shown by electron spin resonance (ESR), nuclear
magnetic resonance (NMR), and infrared (IR) spectroscopy.\cite{long98,long99,long07}

Theoretically, the dynamic JT effect for singly charged fullerene
\C60a is expected to be reflected in the IR spectrum. At room
temperature, the dynamic disorder between different orientations of
the \C60a is signalled in the infrared spectrum by the
splitting of the T$_{1u}$ fundamental modes of fullerene and the activation of
silent modes. Neutral \full\ possesses the highest I$_{h}$ symmetry with 174 possible vibration modes,
out of which only four T$_{1u}$ modes are infrared active due to symmetry considerations.
The temperature dependence of the JT dynamics was
studied on these compounds by far-infrared spectroscopy.\cite{long98,long99,long07,Bietsch2000, Gotschy94, Becker93, Volkel95, Gotschy96}

Another experimental evidence for the dynamic nature of the JT effect in [\PhC60]
was obtained by ESR,\cite{Bietsch2000, Gotschy94, Becker93, Volkel95, Gotschy96}
namely by the splitting of the LUMO of \C60a above 40 K through an Orbach
spin-lattice relaxation process.\cite{Volkel95} The ESR results \cite{Bietsch2000} on
[\textit{A}$^{+}$(C$_{6}$H$_{5}$)$_{4}$]$_{2}$C$_{60}$$^{-}$\textit{B}$^{-}$ (where \textit{A}
= P or As and \textit{B} = I or Cl) indicate static disorder with random occupation of two
``standard orientations'' at low temperature.
Ordering phenomena were also found in x-ray diffuse scattering and diffraction measurements:
Launois et al. \cite{Launois2000} found evidence for a structural phase transition in
(Ph$_{4}$P$^{+}$)$_{2}$C$_{60}$$^{-}$Br$^{-}$ at around 120 K during cooling down,
where the C$_{60}$ molecules show an orientational order with the formation of two types of
orientational domains. Simultaneously, the average crystal structure is changed from I4/m to I2/m,
and the related lowering of the crystal field symmetry leads to a static stabilization of
the JT distortion of the \C60a molecule.\cite{Launois2000} This scenario is consistent with
the results of far-infrared transmission measurements, which revealed a weak transition in
the range 125-150~K.\cite{long98}
Later on, based on x-ray diffraction data\cite{Pilz02} a glass transition was proposed to occur in
Ph$_{4}$As)$_{2}$ClC$_{60}$ at 125~K, where the dynamic disorder of the \full\ molecules
over two orientations becomes static. Besides, an upper bound of 0.01\AA\ for the
JT distortion of the \full\ molecules was established.\cite{Pilz02}

\begin{figure}[t]
\begin{center}
\includegraphics*[scale=0.6]{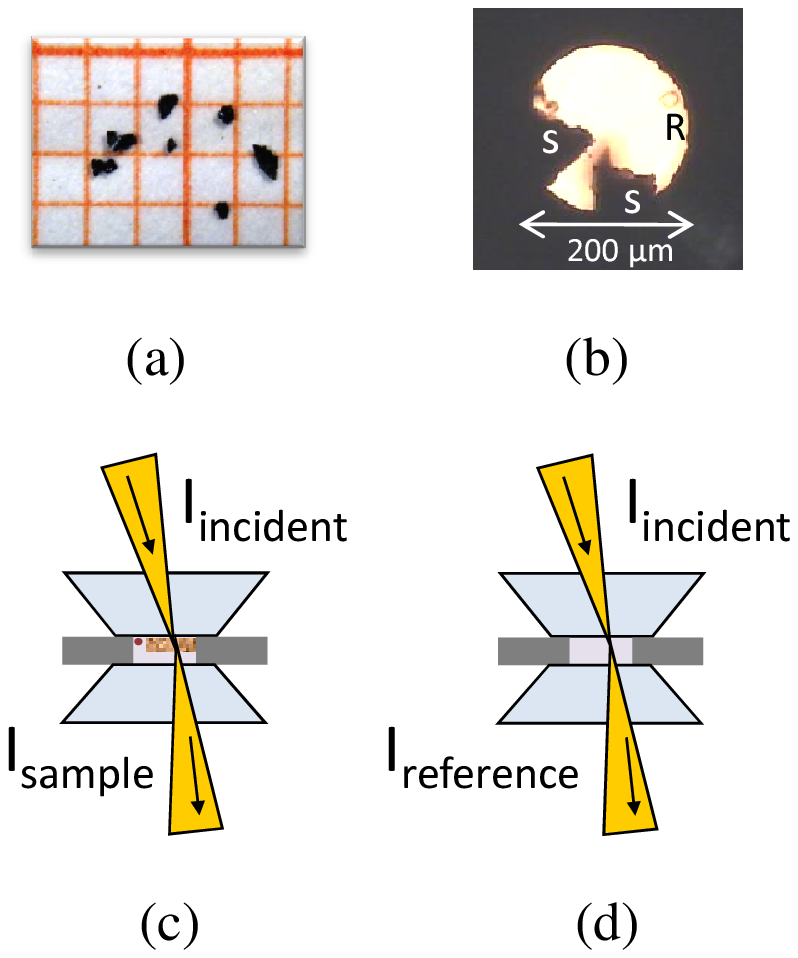}
\caption{(a) Photo of \PhC60 single crystals on millimeter
paper. (b) Microscopic view of the DAC filled with samples (S), ruby
ball (R), and pressure transmitting medium (bright area). Scheme of (c) the
transmission through the sample, and (d) the transmission through the
pressure transmitting medium (CsI).} \label{cell}
\end{center}
\end{figure}

The goal of this work was to study the pressure-induced effects in \PhC60 in detail, as compared
to earlier investigations,\cite{francis10} and to compare them to those
induced by temperature decrease. The interesting similarity between the temperature lowering and
increasing pressure is exhibited by \full~ and several fullerene based compounds
\cite{Sundqvist99,Thirunavukkuarasu07,Thirunavukkuarasu08}. In case of \PhC60\,
temperature lowering causes a decrease of the lattice constants and
the decreasing thermal energy induces a dynamic-to-static JT transition.
In comparison, generally the increase in pressure reduces lattice constants but does
not change the thermal energy of the system. Hence, the changes of the nature of the JT effect
do not necessarily need to be similar during temperature lowering and pressure increase.
The goal of this infrared study was to compare the effects of pressure \textit{versus}
temperature lowering on the vibration properties and the electronic and vibronic
excitations in \PhC60.

We present the results of high-pressure infrared studies on
\PhC60 up
to 9 GPa over a broad frequency range between 300--20000 \cm-1. As the time scale of infrared measurements
(10$^{-11}$~s$^{-1}$) is of the order of the JT pseudorotation frequency of
the \C60a ion, IR spectroscopy is a powerful tool to investigate the
JT dynamics. We can address the symmetry changes of the molecule as
a function of pressure.
Investigating the electronic transitions of the \C60a anion can
help to understand the intermolecular interactions, charge
transfer process, and electronic states.
For example, it was proposed that the halogen anion radius changes the charge distribution
on \C60a, which may alter its vibrational characteristics and electronic absorption.\cite{Semkin}
The pressure dependence of the vibrational and electronic
properties of \PhC60 are thus studied in detail.

\section{Experiment}
\subsection{Synthesis of (Ph$_{4}$P)$_{2}$IC$_{60}$ crystals}
Small single crystals of (Ph$_{4}$P)$_{2}$IC$_{60}$ were grown by
electro-crystallization over platinum cathode with a constant
current of 30 $\mu$A. Electrolysis was carried out in the solution of
Ph$_{4}$PI and C$_{60}$ dissolved in 1:1 mixture of dichloromethane
and toluene at ambient conditions \cite{Allemand91}. Black shiny
crystals of 100-300 $\mu$m size [see Fig. \ref{cell} (a)]
were collected from the electrode after 5 days.

\begin{figure}[t]
\begin{center}
\includegraphics*[scale=0.8]{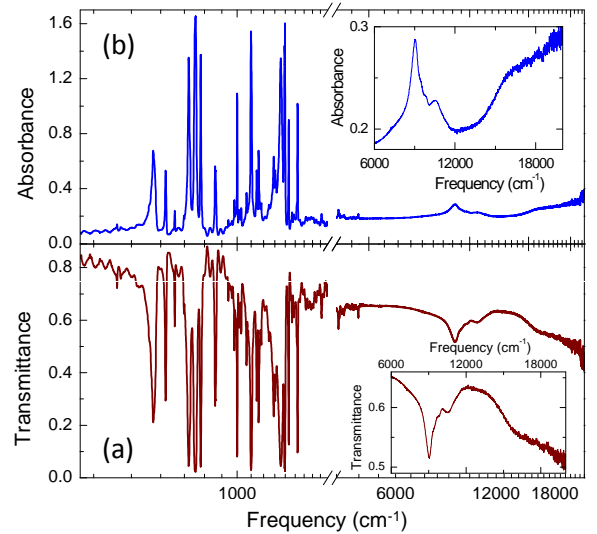}
\caption{(a) Infrared transmittance and (b) corresponding absorbance spectra of \PhC60
in the frequency range 300--20000 \cm-1 at 0.6~GPa. Insets: Transmittance and
absorbance spectra in the NIR-VIS region.} \label{fullrange}
\end{center}
\end{figure}

\subsection{High-pressure infrared measurements}

Infrared transmission measurements were carried out with an infrared
microscope Bruker IR scope II with 15x magnification coupled to a
Bruker 66v/S Fourier transform infrared spectrometer. The high
pressure was generated by a Syassen-Holzapfel diamond anvil cell
(DAC) \cite{Huber77} equipped with
type IIA diamonds suitable for infrared measurements. The ruby luminescence method was used for pressure
determination \cite{Mao86}. The transmission was measured for
pressures up to 9 GPa  between 200 \cm-1
and 20000 cm$^{-1}$. Finely ground CsI was used as quasi-hydrostatic
pressure transmitting medium. Data were collected with resolution of
1 cm$^{-1}$ for 100--600 cm$^{-1}$, 2 cm$^{-1}$ for 550--8000 cm$^{-1}$
frequency range. For the transmission measurements in the NIR-VIS region
a powder sample was mixed with CsI and filled in the DAC.
Measurements were carried out with 4~cm$^{-1}$ resolution in \textbf{the} NIR-VIS region.
All measurements were carried out at room
temperature. A microscopic view of the DAC filled with samples,
the pressure transmitting medium, and the ruby ball is shown in Fig.~
\ref{cell}(b).

\begin{figure}[t]
\begin{center}
\includegraphics*[scale=0.7]{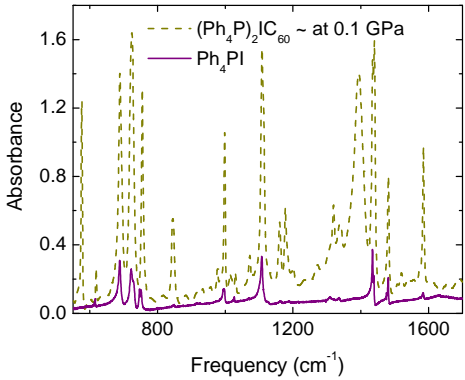}
\caption{Infrared absorbance spectrum of \PhC60 at $\sim$~0.1 GPa and of pure
(Ph$_{4}$P)$_{2}$I at ambient pressure.} \label{compspectra}
\end{center}
\end{figure}

\begin{figure}[b]
\begin{center}
\includegraphics[scale=0.6]{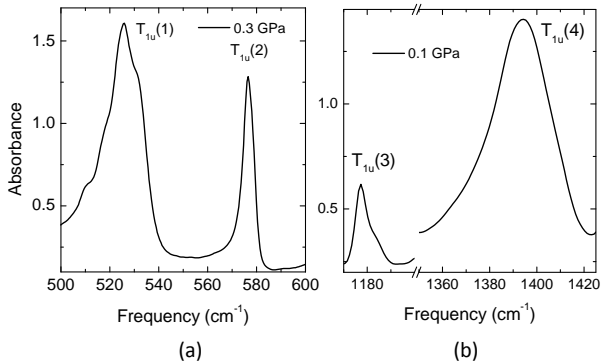}\\
  \caption{T$_{1u}$ modes of \C60a at the lowest applied pressure.}\label{funda}
  \end{center}
\end{figure}

In order to determine the transmittance of \PhC60 under pressure,
the intensity \textit{I}$_{\rm s}$($\omega$) of the radiation transmitted by the sample
or by the mixture of the powder sample and the pressure transmitting medium
was measured, as illustrated in Fig.~\ref{cell}(c). As reference, the
intensity \textit{I}$_{\rm r}$($\omega$) transmitted by the pressure
transmitting medium inside the DAC was used, as shown in
Fig.~\ref{cell}(d). The transmittance was then calculated according to
\textit{T}($\omega$)=\textit{I}$_{\rm s}$($\omega$)/\textit{I}$_{\rm r}$($\omega$)
and the absorbance is given by $\textit{A}$=log$_{10}$(1/$\textit{T}$).

\section{RESULTS AND DISCUSSION}\label{results}

\subsection{Assignment of excitations at lowest pressure} \label{assignment}

\begin{figure}[t]
\begin{center}
\includegraphics*[scale=0.8]{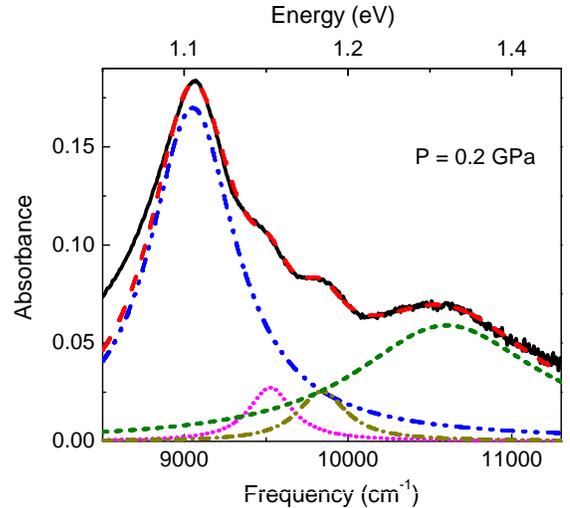}
\caption{Absorbance spectrum of \PhC60 at 0.2 GPa in the NIR region together with the fit (red dashed line) and its Lorentz oscillator contributions. } \label{nir_fit}
\end{center}
\end{figure}

\begin{table*}
\caption{\label{allmode} Vibrational modes of \PhC60\ with their pressure
dependence and assignment. The strength of the modes is specified
as strong (s), medium (m), and weak (w)}
\begin{ruledtabular}
\begin{tabular}{lccr}
Mode position & Pressure dependence & Strength & Assignment\\
(at $\sim$ 0.1 GPa)& & & \\
(cm$^{-1}$)& & & \\
\hline
398 & hardens, doublet above 2 GPa & w &G$_{u}$(1)\\
509 & softens & w & -\\
525 & hardens & s &cation\\
517, 533 & doublet, hardens & m,s & T$_{1u}$(1)\\
576, 578 & doublet, hardens & s,m & T$_{1u}$(2)\\
619 & hardens, doublet above 2 GPa & w &cation\\
665 & sharpens, hardens, doublet $\sim$ 2 GPa & w & -\\
689, 695 & doublet, hardens & s,m & cation\\
721, 726 & doublet, hardens, triplet above 2GPa & s,s & cation\\
756 & slope change above 2 GPa & s & cation\\
844, 848 & doublet, hardens & w,w & \C60a\\
975 & gains intensity, hardens & w & cation\\
998 & doublet above 2 GPa, hardens & m & cation\\
1072 & hardens & w& \C60a\\
1108, 1114 & doublet, hardens & s,s &-\\
1177, 1182 & doublet, hardens & w,w & T$_{1u}$(3)\\
1201 & hardens, undetectable above 3 GPa & w & -\\
1320 & hardens, doublet at very high P & w &cation\\
1364, 1395 & doublet at low P; hardens & w,s
 & T$_{1u}$(4)\\
1435, 1440 & hardens, doublet & s,s & cation\\
1482 & hardens, doublet above 2 GPa & m & cation\\
1585 & hardens, sharp up to high P & m & cation\\
3047, 3056, 3076, 3088& multiplet, hardens & m,s,w,w& -\\
\end{tabular}
\end{ruledtabular}
\end{table*}

\PhC60 contains singly charged fulleride anions,
with nearly isolated \C60a anions in the solid state environment due to the large sized cation (see Fig.~\ref{crystal}).
The infrared transmittance and absorbance spectra of \PhC60 are
presented in Fig.~\ref{fullrange} between 300 and 20000 \cm-1 at
room temperature. A small region of the spectrum is cut out between 2000-4000 \cm-1 due to multiphonon absorption in the diamond anvils. The vibrational modes are observed in the far- and mid-infrared region, whereas the electronic transitions appear in the NIR-VIS region between 6000--12000 \cm-1 (see inset of Fig. \ref{fullrange}). The frequency positions, relative strengths, and assignments of the vibrational modes are listed in Table \ref{allmode}.

It is apparent from the infrared spectrum of \PhC60, that it contains numerous vibrational modes of \C60a\ and of the
(Ph$_{4}$P$^{+}$)$_{2}$I cation in the FIR and MIR region. In order to illustrate the contribution of the cation to the richness of the \PhC60  vibrational spectrum, Fig.~\ref{compspectra} shows the infrared absorbance spectrum of \PhC60 (at 0.1 GPa) in comparison with that of pure (Ph$_{4}$P)$_{2}$I at ambient conditions.

\begin{figure}[t]
 \begin{center}
\includegraphics*[scale=0.5]{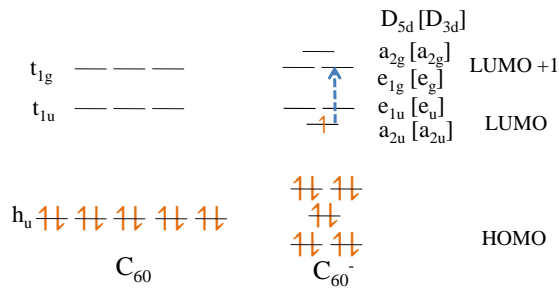}\\
  \caption{Illustration of molecular orbitals of \full\ and \C60a for D$_{5d}$[D$_{3d}$] symmetry according to Refs.\ \onlinecite {Green96,Lawson1992}.
  The blue dotted line indicates the optically allowed transition in \C60a.}\label{scheme}
  \end{center}
\end{figure}

We first focus on the four fundamental T$_{1u}$ modes of \C60a\ and the electronic
transitions observed in the lowest-pressure absorbance spectrum of \PhC60, as depicted in Fig.~\ref{funda}
and \ref{nir_fit}, respectively. Neutral \full\
has a triply degenerate, empty LUMO (lowest unoccupied molecular orbital) and a completely filled
HOMO (highest occupied molecular orbital) (see Fig.\ \ref{scheme}). When the \full\
molecule is doped with electrons, the symmetry is lowered depending on the
number of electrons added. The LUMO of \full\ can be occupied by up to six electrons. Such addition of electrons to the \full\ molecule causes a
disturbance in the spherical distribution of the electron cloud. In
case of \C60a, the additional electron causes a change in the C--C and
C=C bonds near the poles. Such stretched bonds are nearly in the
direction of the symmetry axis, therefore the spherical \full\
becomes ellipsoidal \C60a \cite{Koga92}. This in turn induces the
JT distortion, causing the splitting of the LUMO levels. The JT
effect depends on the number of charges added to the \full. It can induce new electronic transitions,
shifts and splittings of the T$_{1u}$ modes, and can lead to the activation of new modes in the
vibrational spectra.

The T$_{1u}$ modes of \full\ are governed by the electron-phonon coupling depending on the
charge added to the \full\ molecules.  At room temperature the T$_{1u}$ vibrational modes in neutral C$_{60}$ are sharp singlets resonating at 527, 576, 1182 and 1428 cm$^{-1}$,\cite{Kratschmer90} whereas the T$_{1u}$ modes of \C60a in \PhC60 are doublets with the frequencies (1) 517, 533 ,(2) 576, 578, (3) 1177, 1182, and (4) 1364, 1395~cm$^{-1}$. At the lowest measured pressure
all four T$_{1u}$ modes are split into doublets (see Fig. \ref{funda}). All the T$_{1u}$ modes except T$_{1u}$(2) show a redshift compared to \full,\cite{Pichler94} attributed to the coupling of the vibrational mode to virtual
t$_{1u}$$\rightarrow$t$_{1g}$ transitions\cite{Rice92} (see scheme in Fig.\ \ref{scheme}).
Furthermore, the T$_{1u}$ modes show strong enhancement of the line
width and oscillator strength, and also a change in line shape. Doublet splitting of the T$_{1u}$ modes of the
\C60a anion is the signature of the JT effect in
the molecule. The room temperature dynamic JT distortion in \PhC60
was also reported by FIR studies \cite{long98,long99}. Among these four fundamental vibrational modes, the T$_{1u}$(4) mode shows the strongest redshift \cite{Semkin} compared to the neutral \full.  Furthermore there are silent fullerene modes which become infrared active in \PhC60 due to symmetry lowering, like the G$_u$(1) mode at 398~cm$^{-1}$. The infrared-active cation phonon modes contribute to the richness of the absorbance spectrum as well.

The doublet splitting of the T$_{1u}$(1) and T$_{1u}$(2) modes found in our lowest-pressure data is consistent with an earlier report.\cite{long98} Long \textit{et al.} \cite{long98} studied the temperature dependence of vibrational modes and found anomalies
in the shift of the frequency positions in the temperature range 125 -- 150 K. For example, the G$_u$(1) mode at 398 \cm-1 is reported as a singlet at room temperature and undergoes a doublet splitting during cooling below 150 K.
The vibrational mode observed by Long et al. \cite{long98} at 504 \cm-1, which softens on lowering the temperature, is shifted to 509 \cm-1 in our data. Besides, the vibrational mode at 619 \cm-1 only shows a minute frequency shift on lowering the temperature and flattens out below 125 K. The temperature dependence of the vibrational modes will be compared to our pressure-dependent results presented in Section \ref{section:pressure dependence}.

In \C60a the electronic states are coupled to vibrational modes, which
gives rise to vibronic transitions. The electronic and vibronic
transitions are clearly observed between 6000-12000 \cm-1 in the NIR
region in Fig. \ref{fullrange}. For better lucidity the NIR-VIS  region is
presented in the inset.
Fig \ref {nir_fit} shows the electronic transition in the NIR-VIS region of the spectrum with the four contributions obtained from the fitting with Lorentz oscillators. To obtain a very good fit on the low-frequency side an additional oscillator is required for describing the background, which does not affect the frequency
positions of the other main oscillators representing the electronic transitions.
Besides the prominent feature at $\sim$9050~\cm-1, three absorption bands between 9300 and 11000~\cm-1 are observed.

\begin{figure}[t]
\begin{center}
\includegraphics*[scale=0.7]{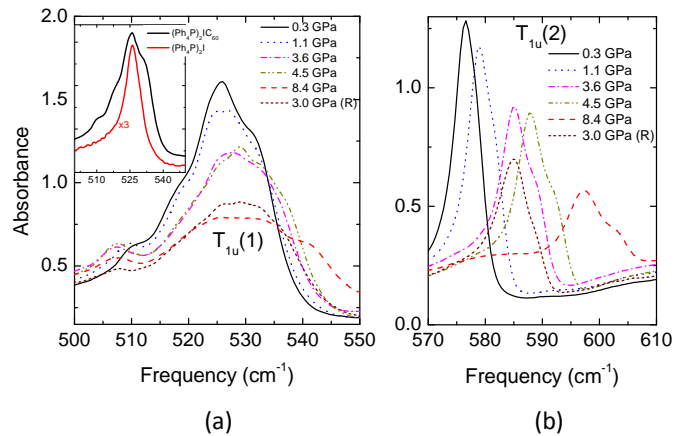}
\caption{(a) T$_{1u}$(1) and (b) T$_{1u}$(2) infrared absorbance spectra of \PhC60 for various pressures. Inset: Comparison of the absorbance spectra of \PhC60 and (Ph$_{4}$P)$_{2}$I between 500-550 \cm-1. The label (R) indicates the spectra measured during pressure release.}
\label{T1u_1}
\end{center}
\end{figure}

\begin{figure}[b]
\begin{center}
\includegraphics*[scale=0.8]{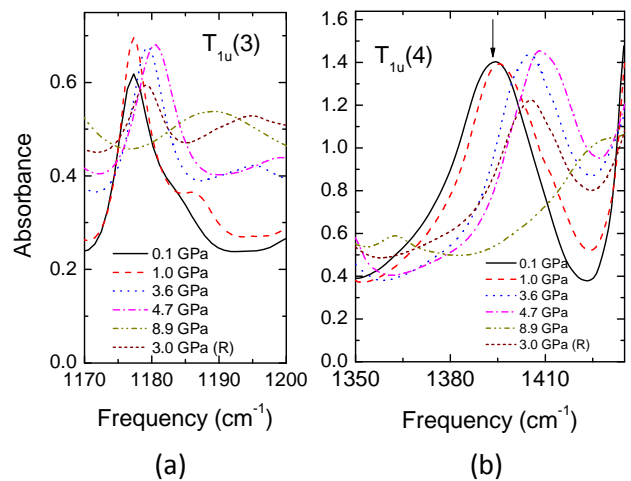}
\caption{ (a) T$_{1u}$(3) and (b) T$_{1u}$(4) infrared absorbance spectrum of \PhC60. The T$_{1u}$(4) mode is marked with an
arrow. The label (R) indicates the spectra measured during pressure release.}
\label{T1u_2}
\end{center}
\end{figure}

\begin{figure}[t]
\begin{center}
\includegraphics*[scale=0.9]{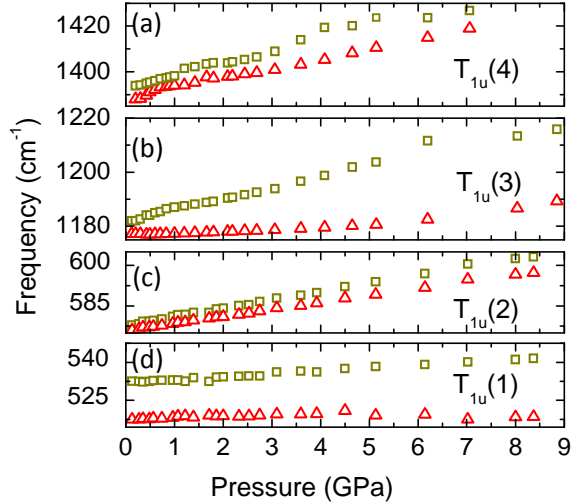}
\caption{Frequency positions of the T$_{1u}$ vibrational
modes as a function of pressure.} \label{pdep}
\end{center}
\end{figure}

\begin{figure}[b]
\begin{center}
\includegraphics*[scale=1]{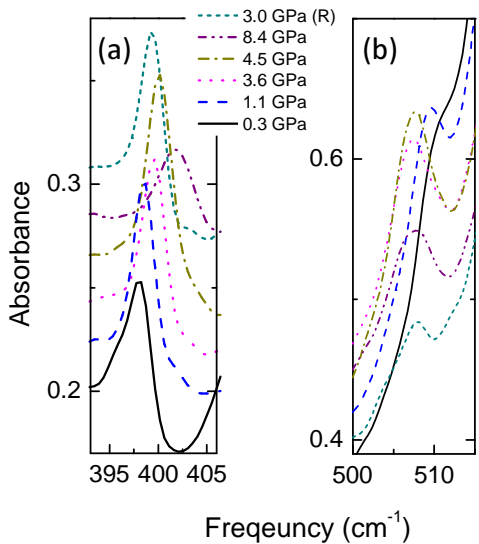}
\caption{Absorbance spectra of the vibrational modes of \PhC60 in the FIR region for various pressures (shifted for clarity).
The label (R) indicates the spectra measured during pressure release.}
\label{firmodes}
\end{center}
\end{figure}

\begin{figure}[t]
\begin{center}
\includegraphics*[scale=0.9]{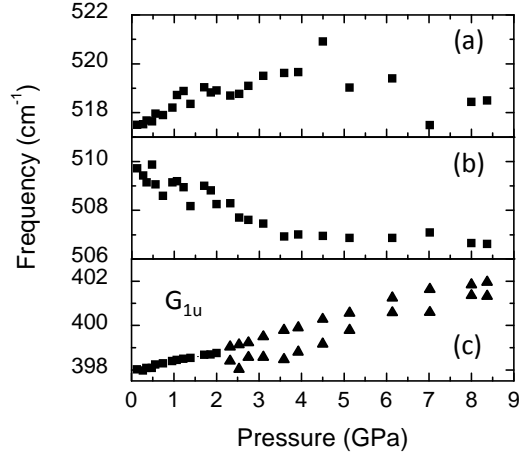}
\caption{Pressure dependence of the frequency positions of the cation modes close to the T$_{1u}$(1) mode [(a) and (b)], and of
the G$_{1u}$ vibrational mode (c).} \label{pdepfirmode}
\end{center}
\end{figure}

Several NIR investigations report similar spectra on \C60a
in solution \cite{Kondo95} and isolated \C60a \cite{Tomita05}.
Electronic transitions of isolated \C60a\ in neon matrices showed well resolved spectra.\cite{Fulara1993} The environment in which the \C60a anion is investigated is an important criterion to
determine the nature of the distortion which can either be static or dynamic. On the one hand, \full\ anions in the solid state are influenced by Coulomb interactions with the cations.  On the other hand, even in dilute solutions the influence of the environment can be significant.

In the following, we discuss the four distinct transitions in the NIR-VIS region of
the spectrum with the help of the transition
scheme in Fig. \ref{scheme}. This scheme is based on the splitting of the t$_{1u}$ and t$_{1g}$ LUMO energy levels, whose degeneracy is lifted in \C60a because of the JT distortion and the related symmetry lowering. According to theoretical investigations, the symmetry for \C60a gives rise to the three possible point groups D$_{5d}$, D$_{3d}$ or
D$_{2h}$, which possess nearly the same JT energies.\cite{Koga92} On an adiabatic potential energy surface there are 6 equivalent
structures possible for D$_{5d}$ minima, 10 for D$_{3d}$ minima,
and 15 for D$_{2h}$ minima.\cite{Koga92} Due to the equivalent
energy the dynamic transformation among these distortions can take
place. In case of D$_{5d}$ symmetry, the orbitals ($^{2}$t$_{1u}$ and $^{2}$t$_{1g}$) undergo a doublet splitting
into ($^{2}$a$_{2u}$, $^{2}$e$_{1u}$) and ($^{2}$e$_{1g}$, $^{2}$a$_{2g}$), respectively, while
for D$_{3d}$ it would be
($^{2}$a$_{2u}$, $^{2}$e$_{u}$) and ($^{2}$e$_{g}$, $^{2}$a$_{2g}$),
respectively. The splitting of the energy levels is illustrated in Fig. \ref{scheme}.
For both D$_{5d}$ and D$_{3d}$ symmetry reduction a single optically
allowed transition of the form a$_{2u}$ $\rightarrow$
e$_{g}$ is expected (see Fig. \ref{scheme}). The different orientations for D$_{5d}$ and D$_{3d}$
distortions are separated by shallow energy minima which are connected by pseudorotation causing the disorder in
the dynamic system.
In case of D$_{2h}$ symmetry, the $^{2}$t$_{1u}$ and
$^{2}$t$_{1g}$ levels undergo a triplet splitting into ($^{2}$b$_{1u}$,
$^{2}$b$_{2u}$, $^{2}$b$_{3u}$) and ($^{2}$b$_{1g}$, $^{2}$b$_{2g}$,
$^{2}$b$_{3g}$), respectively. Thus, a D$_{2h}$ distortion of the fullerene molecule would give rise
to two optically allowed transitions of the type b$_{1u}$ $\rightarrow$ b$_{3g}$
and b$_{1u}$ $\rightarrow$ b$_{2g}$ \cite{Kondo95}. It has been suggested that the D$_{2h}$ symmetry can be
stabilized only in a crystal field, but not in the case of a free
\C60a\cite{Chaney97}. The possibility of D$_{2h}$ symmetry for \C60a has been
ruled out due to the narrow line in the FIR spectrum
reported earlier \cite{long98}.

\begin{figure*}[t]
\begin{center}
\includegraphics*[scale=0.8, angle=-90]{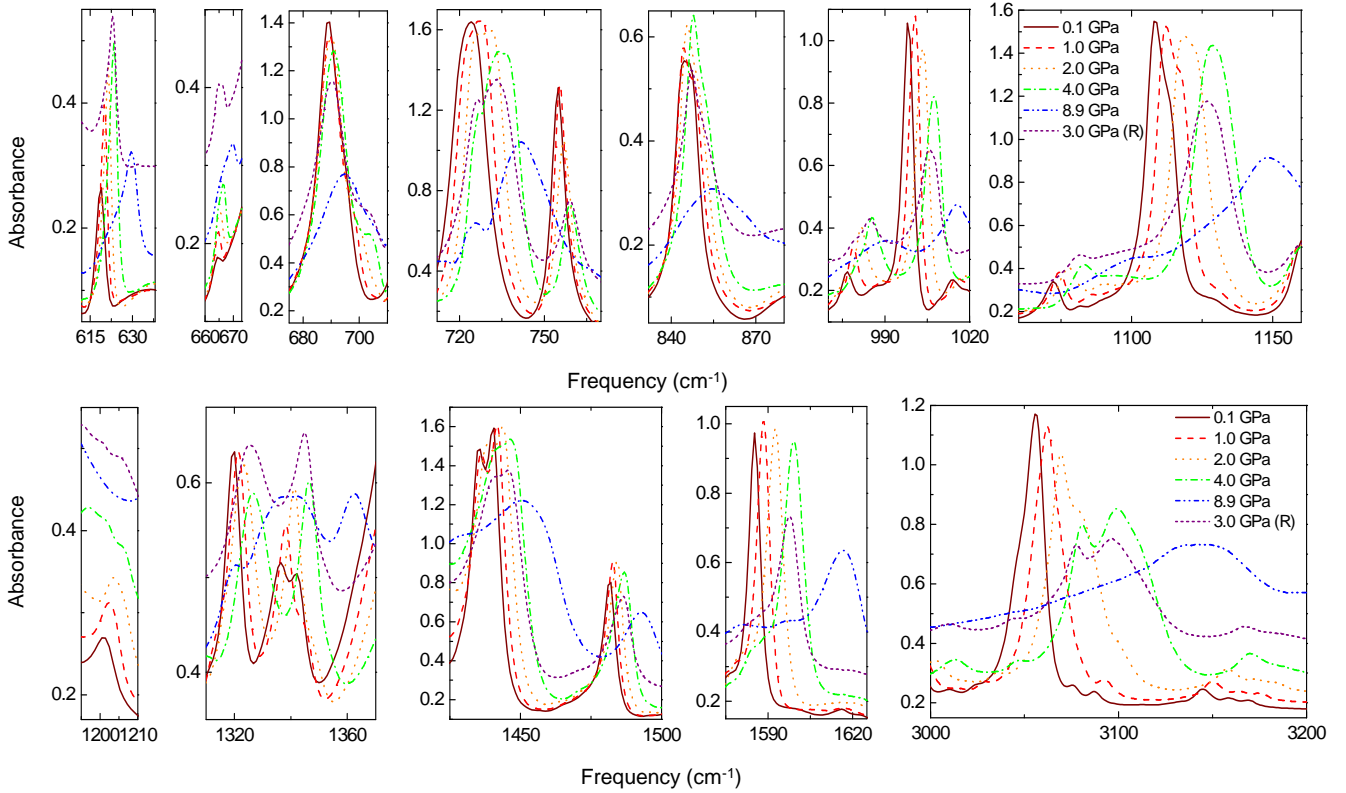}
\caption{Absorbance spectra of the vibrational modes of \PhC60 in the MIR region for various pressures. The label (R) indicates the spectra measured during pressure release.}
\label{mir}
\end{center}
\end{figure*}

\begin{figure}[h]
\begin{center}
\includegraphics*[scale=0.8]{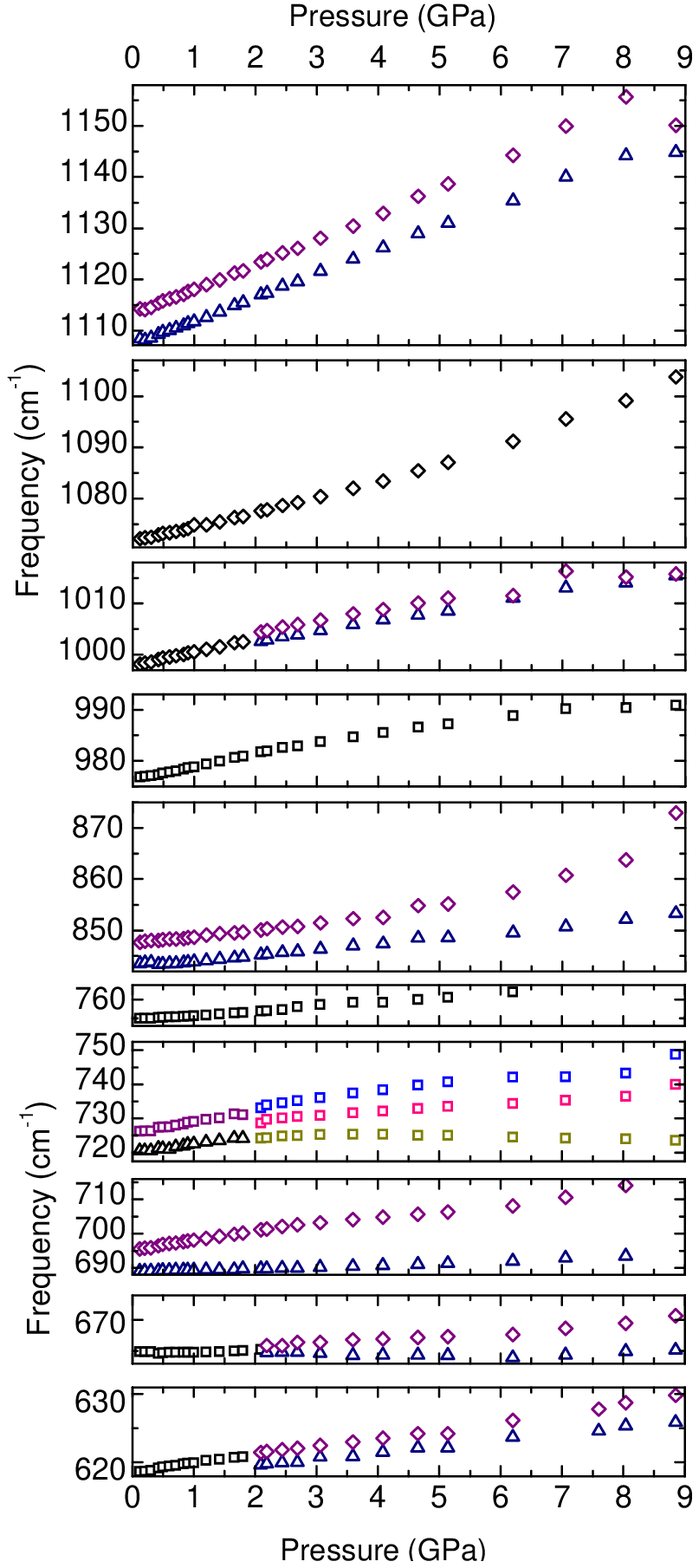}
\caption{Frequency positions of the vibrational modes of \PhC60 as a function of pressure.}
\label{mirpdep_1}
\end{center}
\end{figure}

\begin{figure}
\begin{center}
\includegraphics*[scale=0.8]{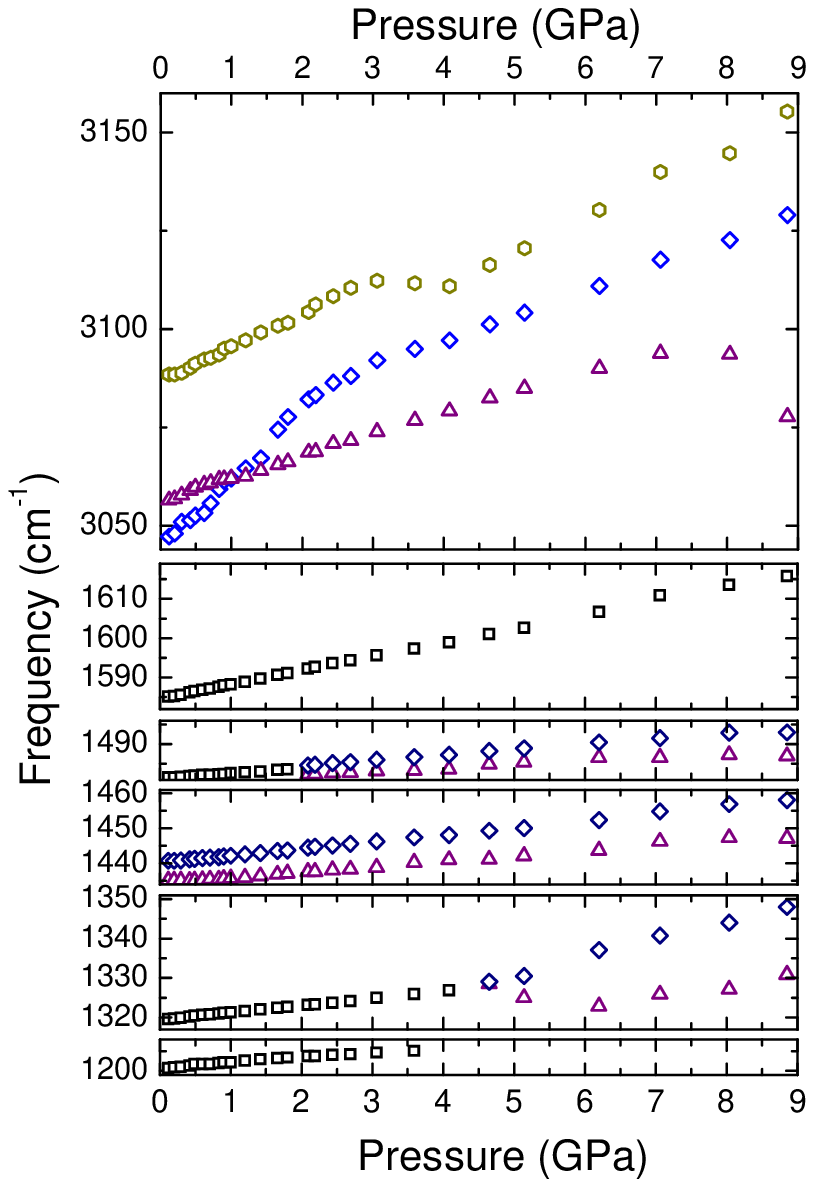}
\caption{Frequency positions of various vibrational modes of \PhC60 as a function of pressure.}
\label{mirpdep_2}
\end{center}
\end{figure}

According to Lawson et al. \cite{Lawson1992} the features in the NIR region are due to the symmetry reduction to D$_{5d}$ of the isolated \C60a anion investigated in benzonitrile solution. The strong feature at 1078 nm (9276 \cm-1) in the spectrum is attributed to the optically allowed a$_{2u}$ $\rightarrow$ e$_{1g}$ transition which is in accordance with the density functional calculations by Green
et al. \cite{Green96}. The manifold around 800--1000 nm (10000--12500 \cm-1), which is not very well resolved in Ref.\ \onlinecite{Lawson1992}, is assigned to the vibronic transitions to the level a$_{2g}$. In a recent NIR investigation on \C60a carried out by Hands et al. \cite{Hands08}, a well-resolved spectrum in the region 9000--13000 \cm-1 is presented. They discuss the possibility of D$_{3d}$ and D$_{5d}$ symmetry and claim that the four contributions in the NIR region are due to the D$_{3d}$ symmetry, and that the spectrum would have fewer contributions in case of D$_{5d}$ symmetry. \C60a ion prepared in other media like in gas matrix \cite{Fulara1993},
by electro-generation \cite{G.A.Heath} and in salts \cite{Bolskar95}
were also studied at ambient conditions. Also theoretical
calculations \cite {Koga92,I.D.Hands} have been carried out to explain
the complicated electronic transition observed in \C60a. Hands et al.\cite{I.D.Hands} state that the dynamics
for a minimum of D$_{5d}$ symmetry is simpler than that of D$_{3d}$ due to the tunneling splitting between symmetry adapted states that correctly describe tunneling between equivalent minima. Obviously, there are alternative explanations for the results reported in Ref.\ \onlinecite{Lawson1992}.

According to the above described earlier theoretical and experimental results, we interpret our NIR absorbance spectrum for the lowest pressure in terms of a \C60a molecule dynamically fluctuating between D$_{3d}$ and D$_{5d}$ symmetry. Within this picture, the t$_{1u}$ and
t$_{1g}$ (LUMO and LUMO+1, respectively) levels are split into two levels \cite{I.D.Hands, Green96} (see scheme in Fig. \ref{scheme}). We interpret the NIR spectrum as the combination of electronic and manifold vibronic transitions. The prominent feature at 9049.8~\cm-1 is due to the optically allowed a$_{2u}$$\rightarrow$e$_{1g}$ transition either in D$_{5d}$ or D$_{3d}$ symmetry. The manifold between 9300 and 11000~\cm-1 is due to vibronic transitions. Based on our FIR-MIR data and earlier results\cite{long98} we suggest the symmetry of \C60a near ambient conditions to be governed by the dynamic JT effect (presumably of D$_{3d}$ and D$_{5d}$ symmetry). However, a further symmetry lowering, for example from D$_{5d}$ to C$_{2h}$ or C$_i$, cannot be ruled out, as pointed out recently in the case of (Ph$_4$As)$_2$ClC$_{60}$.\cite{long07}

\subsection{Pressure dependence of vibrational modes and electronic transitions of \PhC60} \label{section:pressure dependence}

Next we will focus on the effect of pressure on the vibrational and electronic excitations in \PhC60.  For a quantitative analysis, the frequency positions of the
vibrational modes were extracted by fitting the modes with
Lorentzian functions. The pressure dependence of all modes is summarized in Table \ref{allmode}. Fig. \ref{T1u_1} and \ref{T1u_2} show the four fundamental T$_{1u}$ vibrational modes of \C60a for selected
pressures. The T$_{1u}$(1) is strongly overlapped by a
counterion mode, as illustrated in the inset of Fig. \ref{T1u_1} (a). Also the (Ph$_{4}$P)$^{+}$ cation modes undergo pressure-induced changes [see Fig. \ref{T1u_1} (a)]; this complicates the analysis of this mode. The pressure dependent frequencies of T$_{1u}$ vibrational
modes are plotted in Fig. \ref{pdep}. It is evident that all the T$_{1u}$ modes
are split into doublets at near-ambient conditions and harden with increasing
pressure. We do not observe any anomaly in the pressure dependence of their frequency positions, in contrast to the findings as a function of temperature \cite{long98}. Fig. \ref{pdepfirmode} (a) shows the pressure
dependence of the cation modes at around 518 \cm-1, close to the T$_{1u}$(1) mode's position.

The other prominent vibrational modes in the FIR region are the G$_{u}$(1) mode at 398~\cm-1, depicted in Fig. \ref{firmodes} (a) for various pressures, and another mode at 509~\cm-1 [see Fig. \ref{firmodes}~(b)], whose assignment is not clear. The frequency positions of these modes are plotted as a function of pressure in Figs.\ref{pdepfirmode}~(b) and (c). The G$_{u}$(1) mode is a singlet at low pressure, hardens with increasing pressure, and becomes a doublet
above 2~GPa [see Fig. \ref{pdepfirmode} (c)]. This behavior is consistent with the temperature-dependent results of \PhC60\, where the G$_{u}$(1) undergoes a two-fold splitting at 150~K while cooling down from room temperature.\cite{long98} According
to group theory, a two-fold splitting of the G$_{u}$ mode is expected as the system settles for a lower symmetry. The vibrational mode at 509~\cm-1 is one of the few
modes which soften with increasing pressure, which is also consistent with the behavior during temperature decrease. The intensity of this mode steadily increases with increasing pressure and remains sharp until the highest
pressure applied.

\begin{figure}[h]
\begin{center}
\includegraphics*[scale=0.8]{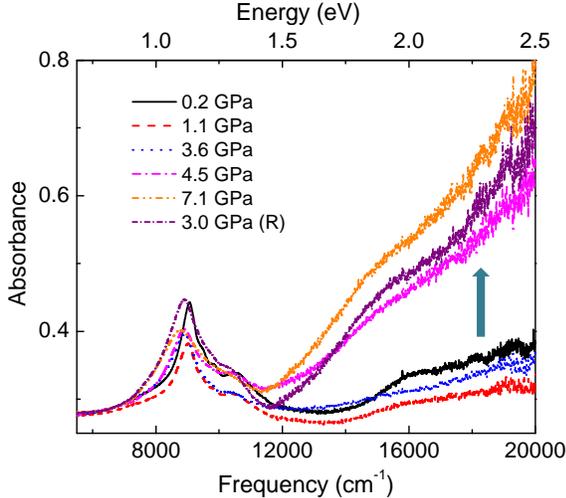}
\caption{Infrared absorbance spectrum of \PhC60 in the NIR-VIS region for various pressures. The label (R) indicates the spectra measured during pressure release.}
\label{nirvis}
\end{center}
\end{figure}

Fig. \ref{mir} shows the evolution of the absorbance spectrum with increasing
pressures for various vibrational modes observed in the mid-infrared
region of the spectrum. There are numerous modes
in this range, and we will discuss the prominent ones in detail in
this section. Their pressure dependence is
included in Table \ref{allmode}. Fig. \ref{mirpdep_1} and \ref
{mirpdep_2} show the pressure-dependent frequency position
of the MIR vibrational modes: All the modes
show a hardening behavior with increasing pressure. The
vibrational mode at 619 \cm-1 is a singlet at the lowest pressure
and undergoes a doublet splitting at pressures above 2~GPa.  In contrast, by lowering the temperature at ambient pressure no splitting of this mode occurs. This
mode is attributed to the cation. The vibrational mode at 665~\cm-1
is attributed to the \C60a \cite{long07}. It is a weak mode but
gains intensity and remains sharp up to high pressure, and also
shows a two-fold splitting on increasing pressure
above 2~GPa. There are several vibrational modes observed due to the cation between 680 and 3150~\cm-1 in the MIR region. The vibrational mode around 720~\cm-1 is a cation mode and is a doublet at the lowest pressure;
above 2 GPa it transforms to a three-fold mode. The two-fold
vibrational mode around 846~\cm-1 might be attributed to the
\C60a vibration \cite{long07}. The vibrational mode at
1482 \cm-1 is a singlet at low pressure and undergoes a two-fold
splitting above 2 GPa. The vibration around 3050~\cm-1 is an intense doublet which is followed by weak modes on the high-energy side at 3076 and 3088 \cm-1.  The influence of pressure on this multiplet has different pressure coefficients. The weak modes on the higher energy side cannot be observed at higher pressures.  In the frequency \textit{versus} pressure plot shown in Fig. \ref{mirpdep_2} there appears to be a crossing over of the vibrational mode around 1 GPa. This is mainly due to different pressure coefficients of the modes; unfortunately, the origin of these modes is not clear.

The observed splitting of several vibrational modes is an indication of a change in symmetry. Several FIR and MIR vibrational modes show a splitting above 2~GPa. This can be understood as the molecule exhibits dynamic distortions at near-ambient conditions with either D$_{5d}$ or D$_{3d}$ symmetry and undergoes a transition to static state with lower symmetry (D$_{3d}$ or lower). The critical pressure of this transition is around 2~GPa. We speculate here that the dynamic-to-static transition induced by external pressure is analogous to the observed transition at around 150~K,\cite{long07} the driving force being the cation-anion interaction (steric crowding).

\begin{figure}[t]
\begin{center}
\includegraphics*[scale=0.8]{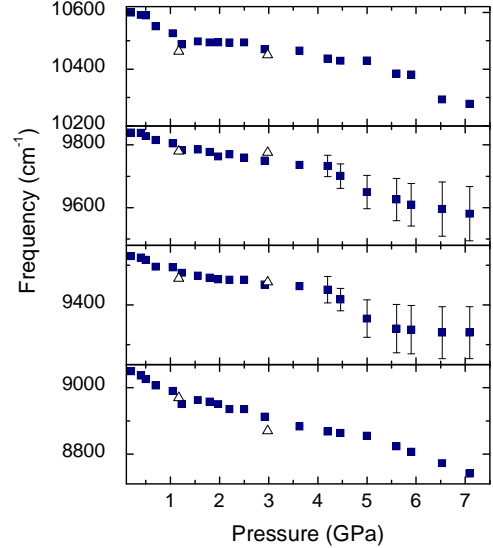}
\caption{Frequency positions of the electronic excitations of \C60a in \PhC60 as a function of pressure. At higher pressures the error bars are enlarged because of the broadening of the transitions. Open triangles indicate the results for releasing pressure.} \label{PvsF_nir}
\end{center}
\end{figure}

The NIR-Vis absorbance spectrum of \PhC60 is shown in Fig.~\ref{nirvis} for
selected pressures. The bands between 9000~\cm-1 and
12000~\cm-1 correspond to electronic and vibronic excitations of the \C60a anion, as discussed in Section \ref{assignment}. With increasing pressure they shift to lower energies; the shifts are reversible upon pressure release. The redshift of the transitions is clearer in Fig.~\ref{PvsF_nir}, where the frequency positions of the bands, as extracted from Lorentz fitting, are plotted as a function of pressure.
Since the feature around 9450~\cm-1 and 9700~\cm-1 broaden considerably at high pressure, the error bar is larger above 4~GPa compared to the lower-pressure regime. The softening of the electronic and vibronic transitions could be due to the fact that the compression of the lattice produced by the applied pressure reduces the splitting of the electronic states. The visible region of the spectrum above 12000 cm$^{-1}$ abruptly increases for pressures above $\approx$4~GPa (see Fig. \ref{nir_fit}). Higher-energy data would be needed in order to clearly trace the details of this change as a function of pressure. The pressure-induced changes on this high-energy transition are irreversible above 5~GPa according to our results.

\section{Summary}

In summary, we have studied the pressure dependence of the vibrational and
electronic/vibronic excitations in \PhC60 by infrared transmission measurements up to 9 GPa over a
broad frequency range (200 - 20000~cm$^{-1}$).
The four fundamental T$_{1u}$ modes of \C60a\
are split into a doublet already at the lowest applied pressure and harden with increasing pressure.
Several cation modes and fullerene-related modes split into a doublet at around 2~GPa, the most prominent
one being the G$_{1u}$ mode of fullerene. We interpret these mode splittings in terms of the transition
from the dynamic to static Jahn-Teller effect.
Four absorption bands are observed in the NIR-VIS frequency
range, which correspond to excitations between t$_{1u}$ and t$_{1g}$ LUMO energy levels, split due to the Jahn-Teller
distortion. The optically allowed a$_{2u}$$\rightarrow$e$_{1g}$ transition (either in D$_{5d}$ or D$_{3d}$ symmetry)
and the three energetically higher-lying vibronic transitions shift to lower energies with increasing pressure,
indicating a reduction of the splitting of the LUMO electronic states under pressure application.

\subsection*{Acknowledgments}
We gratefully acknowledge the financial support by the German
Science Foundation (DFG), Hungarian Academy of Sciences under
cooperation grant (DFG/183) and the Hungarian National Research Fund (OTKA) under grant No. 75813.

\end{document}